\documentclass[a4paper,11pt]{article}
\pdfoutput=1 

\usepackage{jcappub}

\usepackage[T1]{fontenc} 

\title{Wormholes in Randall-Sundrum braneworld}

\author[a]{Ayan Banerjee,}
\affiliation[a]{Astrophysics and Cosmology Research Unit, University of KwaZulu Natal, Private Bag X54001, Durban 4000,
South Africa}
\author[b,c]{P. H. R. S. Moraes,}
\affiliation[b]{Departamento de F\'isica, Instituto Tecnol\'ogico de Aeron\'autica (ITA), 12228-900, S\~ao Jos\'e
dos Campos, SP, Brazil}
\affiliation[c]{UNINA - Universit\`a degli Studi di Napoli Federico II - Dipartamento di Fisica - Napoli I-80126, Italy}
\author[b,d]{R. A. C. Correa,}
\affiliation[d]{SISSA-Scuola Internazionale Superiore di Studi Avanzati, via Bonomea, 265, I-34136 Trieste, Italy}
\author[e]{and G. Ribeiro}
\affiliation[e]{UNESP - Universidade Estadual Paulista "J\'ulio de Mesquita Filho" - Departamento de F\'isica e Qu\'imica, 12516-410, Guaratinguet\'a, S\~ao Paulo, Brazil}

\emailAdd{ayan\_7575@yahoo.co.in}
\emailAdd{moraes.phrs@gmail.com} 
\emailAdd{fis04132@gmail.com} 
\emailAdd{ribeiro.gabriel.fis@hotmail.com} 

\abstract{
Braneworld models were firstly proposed as a great alternative for the hierarcy problem of particle physics, by allowing gravitons, differently from the other interacting bosons, to propagate through an extradimensional space named bulk. As time passed by the braneworld setup has also shown to be able to provide interesting results when applied to cosmology and stellar and gravitational wave astrophysics scenarios. In the present work we will construct Randall-Sundrum II braneworld wormholes whose interior space-time admits conformal motion. We show that for a wide range of positive values of the brane tension, it is possible to fill this wormholes with non-exotic matter, that is, matter obeying the energy conditions, departing from an important theoretical shortcoming of General Relativity wormholes.}

\begin{document}
\maketitle





\section{Introduction}\label{sec:int}

The study of astrophysical objects in higher dimensions has recently attracted considerable interest \cite{burikham/2015,paul/2015,lugones/2017,lugones/2015,visinelli/2018,amir/2018}. Particularly, it has become even more attractive and important after the proposal of the Randall-Sundrum (RS) braneworld models \cite{randall/1999,randall/1999b}. According to those, our four-dimensional (4D) universe is a hypersurface embedded in a 5D space-time, the bulk. Braneworld scenarios were motivated by the possibility of solving the so-called mass hierarchy problem and now have also a significant impact on early universe cosmology (in particular, on the inflationary paradigm) \cite{choudhury/2016,choudhury/2015,lidsey/2006} and even on the late-time universe dynamics \cite{sahni/2002}. Braneworld models contain as a new parameter, the brane tension $\sigma$, which is related to the gravitational stability of the brane. Constraints on the brane tension value were obtained in References \cite{mm/2014,lombriser/2009,bohmer/2008,dabrowski/2004}, for example, from different approaches.

Besides the study of black holes \cite{chamblin/2000,emparan/2000,dadhich/2000}, neutron stars \cite{germani/2001,wiseman/2002} and white dwarfs \cite{garcia-aspeitia/2015}, the brane set up may be also applied to the  wormhole (WH) analysis. WHs are tunnels linking two distant regions in the universe \cite{morris/1988,morris/1988b}. The space-time containing such bridges appears as solutions for the field equations of General Relativity (GR). 

Nevertheless, WH solutions are also obtained in extended or generalized versions of GR. Some examples can be seen for Gauss-Bonnet theory \cite{bhawal/1992}, Brans-Dicke theory \cite{agnese/1995,anchordoqui/1997,he/2002,bhadra/2005}, dilaton gravity \cite{eiroa/2005}, $f(R)$ and $f(R,T)$ theories \cite{bertolami/2012,mcl/2017,ms/2018,sms/2018,ms/2017}, with $R$ and $T$ being the Ricci scalar and trace of the energy-momentum tensor, respectively, and Kaluza-Klein gravity \cite{dzhunushaliev/1999}, among others. 

In fact, one can also find braneworld WHs in the literature. In \cite{bronnikov/2003}, the condition $R=0$ was used for obtaining a large class of metrics that may be treated as WH solutions. In \cite{lobo/2007}, the condition $R=0$ was dropped and the WH matter content was shown to satisfy the null energy condition (NEC). In fact, the main motivation for working with WHs in extended theories of gravity is that those can provide, through their extra degrees of freedom, a possibility for the WH matter content to respect the energy conditions, a feature not attained in GR WHs \cite{morris/1988}. This will be deeply discussed some sections below. Other braneworld WHs are those reported in \cite{wong/2011,wang/2018}. In \cite{wong/2011}, particularly, the authors proposed inflating WHs in braneworld scenarios, while in \cite{wang/2018}, WH space-time configurations were constructed in Dvali-Gabadadze-Porrati brane model.

It is important to remark that although so far WHs have not been detected, attempts to do so are constantly offered in the literature, as it can be checked, for instance, in \cite{shaikh/2017,kuhfittig/2014,li/2014,ohgami/2015,tsukamoto/2012,nandi/2017,shaikh/2018}.

In order to reduce the number of unknowns in the set of WH field equations, an equation of state (EoS) is normally invoked for the WH matter content, i.e., a relation between pressure and density inside these objects. With the same purpose, one can also consider that the space-time inside braneworld WHs admits conformal motion, as we are going to assume here.

In GR, it is important to know the behavior of the metric when moved along curves on a manifold. The conformal Killing operator $\mathcal{L}$ associated with the metric $g$ is the linear mapping from the space $\mathcal{J}(\bf \xi)$ of vector fields on ${\bf \xi}$, 
which is translated by the following relationship
\begin{equation}\label{eq0}
\mathcal{L}_\xi  g_{ik}=\psi  g_{ik}, ~~~\text{where}~~~  \xi \in \mathcal{J}(\bf \xi)
\end{equation}
where $ \psi$ is the conformal factor. The above Eq.(\ref{eq0}) is essentially geometric from two points of view, namely, it is a generalization of self-similarity in hydrodynamics (when $\psi$ is not constant); and it also generalizes the property of the incompressible Schwarzschild interior solution which is conformally flat (see Ref. \cite{Maartens}). Moreover, the conformal symmetry  establish a relation between geometry and matter as well as treated a geometrical EoS for closing the system of equations. Specifically  $ \psi$ is not arbitrary, rather it depends on the CKV as $\psi(x^k) = \frac{1}{4} \xi_{;i}^i $ for Riemannian spaces of four dimensions.

As emphasized in Ref. \cite{Maharaj} conformal symmetries have been studied for spherically symmetric spacetimes without specifying the form of the matter distribution. Such extensions has been studied in three-dimensional manifolds admitting Lorentz metrics \cite{Hall}. An immediate consequence of admitting conformal symmetry stellar model has been proposed for both isotropic and anisotropic fluid spheres \cite{Maharaj1,Maharaj2,Ivanov,Maharaj3,Banerjee}.
Besides this, in braneworld models \cite{Harko} conformally symmetric vacuum solutions of the gravitational field equations have been studied. In recent treatments gravastar solutions within the Mazur-Mottola framework admitting conformal motion were found in  \cite{Usmani,Banerjee1}.

Our intention here is to construct RS braneworld WHs whose space-time admits conformal motion. Our main intention will be to check if the brane tension, for some particular range of values, may allow the obedience of the energy conditions. We organize the present paper as follows. In Section \ref{sec:fes} we construct the braneworld WH field equations. We admit the conformal motion of Killing vectors inside the space-time in Section \ref{sec:cm}. In Section \ref{sec:ir} we derive the WH solutions. The Weyl stresses, namely the scalar $\mathcal{U}$ and the non-local anisotropic pressure $\mathcal{P}_{\mu\nu}$, are completely determined. In Section \ref{sec:ec} we apply the energy conditions to the WH matter content solutions. Our final remarks and conclusions are made in Section \ref{sec:frc}.

\section{Wormhole field equations in Randall-Sundrum braneworld formalism}\label{sec:fes}

In the present paper we will be concerned in obtaining WH solutions in RS II braneworld scenario \cite{randall/1999b}. The concepts of this model have been recently applied to different areas, such as inflation \cite{barros/2016} and black holes \cite{wang/2016}. 

Let us start by writing the modified Einstein field equations on the brane for such a model. They can be written as \cite{Shiromizu}

\begin{equation}\label{whfe1}
G_{\mu\nu}=k^2 T^{\text{eff}}_{\mu\nu},
\end{equation}
where $G_{\mu\nu}$ is the Einstein tensor, $k^2=8\pi G_N$, with $G_N$ being the newtonian gravitational constant, the speed of light $c=1$ and $T^{\text{eff}}_{\mu\nu}$ represents the effective energy-momentum tensor, given by

\begin{equation}\label{whfe2}
T^{\text{eff}}_{\mu\nu}= T_{\mu\nu}+\frac{6}{\sigma}S_{\mu\nu}-\frac{1}{k^2}\mathcal{E}_{\mu\nu}.
\end{equation}
In Eq.(\ref{whfe2}), $T_{\mu\nu}$ is the usual energy-momentum tensor. We have chosen the bulk cosmological constant in such a way that the brane cosmological constant vanishes and $\sigma$ is the brane tension. The high-energy and non-local corrections are, respectively, given by

\begin{equation}
S_{\mu\nu}=\frac{1}{4}\left[\frac{1}{3}TT_{\mu\nu}-T_{\mu\alpha}T^{\alpha}_{\nu}+\frac{1}{6}g_{\mu\nu}
\left(3T_{\alpha\beta}T^{\alpha\beta}-T^2\right)\right],
\end{equation}
where $T=T^{\alpha}_{\alpha}$ is the trace of the energy-momentum tensor, $g_{\mu\nu}$ is the metric and
\begin{equation}\label{whfe4}
k^2 \mathcal{E}_{\mu\nu}=-\frac{6}{\sigma}\left[\mathcal{U}\left(u_{\mu}u_{\nu}+\frac{1}{3}h_{\mu\nu}\right)
+\mathcal{P_{\mu\nu}}+\mathcal{Q}_{\mu}u_{\nu}\right].
\end{equation}
Eq.(\ref{whfe4}) represents a non-local source, arising from the 5D Weyl curvature, with $\mathcal{U}$ being the non-local energy density, $u^{\mu}$ being the $4$-velocity such that
$h_{\mu \nu}= g_{\mu \nu}+ u_{\mu }u_{\nu}$ is the projected tensor, $ \mathcal{P}_{\mu\nu }$ is the non-local anisotropic pressure and $\mathcal{Q}_{\mu }$ is the non-local energy flux.

For a static spherically symmetric matter distribution, which represents our case, $\mathcal{Q}_{\mu } = 0 $ and the non-local anisotropic pressure $ \mathcal{P}_{\mu\nu }$ reads
\begin{equation}
\mathcal{P}_{\mu\nu }=\mathcal{P}\left(r_{\mu }r_{\nu}-\frac{1}{3}h_{\mu\nu } \right),
\end{equation}
where $\mathcal{P}$ is the pressure of the bulk and $r_{\mu}$ is the projected radial vector. 

Eq.(\ref{whfe4}) then becomes

\begin{equation}\label{e}
k^2 \mathcal{E}_{\mu\nu}=-\frac{6}{\sigma}\left[\mathcal{U}u_{\mu}u_{\nu}+\mathcal{P}r_{\mu}r_{\nu}
+\frac{1}{3}h_{\mu\nu}(\mathcal{U}-\mathcal{P})\right].
\end{equation}
We observe from Eq.(\ref{e}) that $\mathcal{E}_{\mu\nu}$ $\rightarrow 0$ as $\sigma^{-1} \rightarrow 0$. Using this limit in Eq.(\ref{whfe2}), we obtain $T ^{\text{eff}}_{\mu\nu} = T _{\mu\nu}$, that is, we recover GR.

We consider for $T_{\mu\nu}$ a perfect fluid energy-momentum tensor, having the explicit form

\begin{equation}\label{emt}
T_{\mu\nu}=\rho u_{\mu}u_{\nu}+p h_{\mu\nu},
\end{equation}
with $\rho$ being the matter-energy density and $p$ the total pressure of the WH.

We also consider for the WH the static spherically symmetric line element in spherical polar coordinates, i.e.,

\begin{equation} \label{eq9}
ds^2=e^{\lambda(r)}dr^2+r^2\left(d\theta^{2}+\sin^{2}\theta d\phi^2\right)- e^{\nu(r)}dt^2,
\end{equation}
with $\lambda(r)$ and $\nu(r)$ being the metric potentials.

The gravitational field equations for the energy-momentum tensor \eqref{emt} and line element (\ref{eq9})  must satisfy the effective 4D equations (\ref{whfe1}). They read \cite{Maarteens,Koyama,Ovalle}

\begin{equation}\label{eq10}
e^{-\lambda}(\lambda'r-1)+1=k^2 \rho^{\text{eff}}r^2,
\end{equation}
\begin{equation}\label{eq11}
e^{-\lambda}(\nu'r+1)-1=k^2 p^{\text{eff}}_rr^2,
\end{equation}
\begin{equation}\label{eq12}
e^{-\lambda}(\nu'-\lambda')\left(\frac{1}{2}\nu'+\nu''+\frac{1}{r}\right)=2k^2 p^{\text{eff}}_t,
\end{equation}
\begin{equation}\label{eq13}
p^{\prime}+\frac{\nu^{\prime}}{2}\left(\rho+p\right)=0,
\end{equation}
where primes denote differentiation with respect to $r$. From Eqs.(\ref{eq11}) and (\ref{eq12}), it is clear that the bulk implies in anisotropy in brane objects. The effective energy density $\rho^{\text{eff}}$, effective radial pressure $p^{\text{eff}}_r$ and effective transverse pressure $p^{\text{eff}}_t$ in the equations above read

\begin{equation} \label{eff}
\rho^{\text{eff}}=\rho\left(\frac{\rho}{2\sigma}+1\right)+\frac{6}{k^4 \sigma}\mathcal{U},
\end{equation}
\begin{equation}
p^{\text{eff}}_r=p+\frac{1}{\sigma}\left[\frac{\rho}{2}\left(\rho+2p \right)+\frac{2}{k^4}(\mathcal{U}+2\mathcal{P})\right],
\end{equation}
\begin{equation}\label{eq16}
p^{\text{eff}}_t=p+\frac{1}{\sigma}\left[\frac{\rho}{2}\left(\rho+2p \right)+\frac{2}{k^4}(\mathcal{U}-\mathcal{P})\right].
\end{equation}


\section{Wormholes with conformal motion}\label{sec:cm}

Applying a systematic approach in order to deduce exact
solutions, we now demand that the interior space-time of the WHs admits conformal motion. This immediately places a restriction on the gravitational behaviour of these objects. Thus, the Eq.(\ref{eq0}) can be expressed as
\begin{equation}\label{whcm1}
\mathcal{L}_\xi g_{ik}=\xi_{i;k}+\xi_{k;i}=\psi g_{ik},
\end{equation}
with $ \xi_{i}=g_{ik}\xi^{k}$. Hence, the Eqs.(\ref{eq9}) and (\ref{whcm1}), give the following relation (see \cite{bohmer/2008} for a detailed discussion) $ \xi^{1}\nu'=\psi $, $ \xi^{4}= C_1$, $ \xi^{1}= \frac{\psi r}{2}$ and $ \xi^{1}\lambda' + 2\xi_{,1}^{1}=\psi$. Here, $1$ and $4$ represent the spatial and temporal coordinates $r$ and $t$ respectively.

In particular, using the above set of equations one can obtain the metric functions with following relationship
\begin{eqnarray}
e^{\nu}&=&C_{2}^{2}r^{2}, \label{eq18}\\
e^{\lambda}&=&\left(\frac{C_3}{\psi}\right)^{2},\label{eq19}\\
\xi^{i}&=&C_{1}\delta_{4}^{i}+ \left(\frac{\psi r}{2}\right)\delta_{1}^{i},\label{eq20}
\end{eqnarray}
where  $C_{1}$, $C_2$ and $C_{3}$ are constants of integration.

Therefore, the field equation (\ref{eq10})-(\ref{eq12}) turns out to be
\begin{eqnarray} \label{eq21}
&&\frac{1}{r^2}\left(1-\frac{\psi ^2}{C_{3}^2}\right)-\frac{2\psi \psi^{\prime}}{rC_{3}^2}=k^2 \rho^{\text{eff}} \label{dark1},
\\ \nonumber \\
&&\frac{1}{r^2}\left(\frac{3\psi ^2}{C_{3}^2}-1\right)=k^2 p^{\text{eff}}_r \label{eq22},
\\ \nonumber \\
&&\left(\frac{\psi ^2}{C_{3}^2 r^2}\right)+\frac{2\psi \psi^{\prime}}{rC_{3}^2}
=k^2 p^{\text{eff}}_t. \label{eq23}
\end{eqnarray}

The above system of equations represent a spherically symmetric matter distribution admitting conformal motion on the brane. Taking Eqs. (\ref{eq22}) and (\ref{eq23}) together it is possible to compute the
extra dimensional effects in terms of the conformal factor. In this vein, we have
\begin{equation} \label{eq24}
\frac{6}{k^2 \sigma }\mathcal{P}=\frac{2\psi ^2}{C_3^2 r^2}-\frac{1}{r^2}-\frac{2\psi  \psi ^{\prime}}{C_3^2 r}.
\end{equation}
Thus, the existence of conformal motions imposes strong constraints on the wormhole geometry. The key point concerning this construction is to find WH in terms of $\psi$. 

\section{Wormhole solutions}\label{sec:ir}

In this section, we will focus our attention on the problem of obtaining a class of exact solutions for $\psi (r)$, $\rho (r)$, $p(r)$, $\mathcal{U}(r)$ and $\mathcal{P}(r)$. In order to solve the Eqs.\eqref{eq21}-\eqref{eq23}, let us impose that $\mathcal{P}(r)$ satisfies the EoS $\mathcal{P}(r)=\omega \mathcal{U}(r)$, with constant $\omega$. This has already been done in the literature, as one can check, for instance, Ref.\cite{castro/2014}. One can, therefore, rewrite Eq.\eqref{eq24} in the following form

\begin{equation}\label{eq25}
\mathcal{U}=\frac{k^{2}\sigma }{6\Omega }\left( \frac{2\psi ^{2}}{C_{3}^{2}}-\frac{1}{%
r^{2}}-\frac{2\psi ^{\prime }\psi }{rC_{3}^{2}}\right).
\end{equation}

Now, using Eqs.\eqref{eff}-\eqref{eq16} into Eqs.\eqref{eq21}-\eqref{eq23}, it is easy to conclude that

\begin{eqnarray}
&&\frac{1}{r^{2}}\left( 1-\frac{\psi ^{2}}{C_{3}^{2}}\right) -\frac{2\psi
\psi ^{\prime }}{rC_{3}^{2}}=k^{2}\left[ \rho \left( \frac{\rho }{2\sigma }%
+1\right) +\frac{6}{k^{4}\sigma }\mathcal{U}\right] ,  \label{eq26} \\
&&  \nonumber \\
&&\frac{1}{r^{2}}\left( \frac{3\psi ^{2}}{C_{3}^{2}}-1\right) =k^{2}\left\{
p+\frac{1}{\sigma }\left[ \frac{\rho }{2}\left( \rho +2p\right) +\frac{2}{%
k^{4}}(1+2\omega )\mathcal{U}\right] \right\} ,  \label{eq27} \\
&&  \nonumber \\
&&\left( \frac{\psi }{C_{3}r}\right) ^{2}+\frac{2\psi \psi ^{\prime }}{%
rC_{3}^{2}}=k^{2}\left\{ p+\frac{1}{\sigma }\left[ \frac{\rho }{2}\left(
\rho +2p\right) +\frac{2}{k^{4}}(1-\omega )\mathcal{U}\right] \right\} .
\label{eq28}
\end{eqnarray}

Substituting Eq.\eqref{eq25} in Eq.\eqref{eq26}, we obtain

\begin{equation}
\left( 1+\frac{1}{\omega }\right) \frac{1}{r^{2}}-\left( 1+\frac{2}{\omega }%
\right) \frac{\psi ^{2}}{r^{2}C_{3}^{2}}-2\left( 1+\frac{1}{\omega }\right) 
\frac{\psi \psi ^{\prime }}{rC_{3}^{2}}=k^{2}\rho \left( \frac{\rho }{%
2\sigma }+1\right) .  \label{eq29}
\end{equation}

In order to reduce the number of functions and looking for analytical solutions, it is both useful and natural to use an EoS to describe the material content of the WH. Here, we consider that $p(r)=\beta \rho(r)$, with constant $\beta$. In this way, the Eqs.\eqref{eq27} and \eqref{eq28} become

\begin{eqnarray}
&&\frac{1}{r^{2}}\left( \frac{3\psi ^{2}}{C_{3}^{2}}-1\right) =\frac{%
(1+2\beta )k^{2}}{2\sigma }\left( \rho +\frac{\beta \sigma }{1+2\beta }%
\right) ^{2}-\frac{\beta ^{2}\sigma k^{2}}{1+2\beta }+\frac{2}{k^{4}\sigma }%
(1+2\omega )\mathcal{U},  \label{eq30} \\
&&  \nonumber \\
&&\left( \frac{\psi }{C_{3}r}\right) ^{2}+\frac{2\psi \psi ^{\prime }}{%
rC_{3}^{2}}=\frac{(1+2\beta )k^{2}}{2\sigma }\left( \rho +\frac{\beta \sigma 
}{1+2\beta }\right) ^{2}-\frac{\beta ^{2}\sigma k^{2}}{1+2\beta }+\frac{2}{%
k^{4}\sigma }(1-\omega )\mathcal{U}.  \label{eq31}
\end{eqnarray}

With the purpose of obtaining analytical solutions, let us assume that the function $\mathcal{U}$ can be modeled as

\begin{equation} \label{eq32}
\mathcal{U}(r)=\mathcal{U}[\rho(r)]=A \rho^{N}(r)+B.
\end{equation}
where $A$ and B are arbitrary constants and $N$ is a positive integer. Hence, from the preceding function it is possible to put  Eq.\eqref{eq29} in the form

\begin{eqnarray}
&&-k^{2}\rho \left( 1+\frac{\rho }{2\sigma }\right) +\frac{1+\omega }{%
r^{2}\omega }-\frac{2+\omega }{\omega }\left[ \frac{1}{3r^{2}}+\frac{\rho
^{N}}{3}+\frac{k(1+2\beta )}{6\sigma }\left( \rho +\frac{\beta \sigma }{%
1+2\beta }\right) ^{2}\right]   \nonumber \\
&&  \label{eq34} \\
&&\left. -\frac{(1+\omega )}{\omega }\left[ -\frac{1}{3r^{2}}-\frac{%
k^{2}\beta ^{2}\sigma }{1+2\beta }+\frac{k^{2}\beta ^{2}\sigma (1-\omega )}{%
(1+2\beta )(1+2\omega )}+\rho ^{N}\left( -\frac{1}{3}+\frac{2(1-\omega )}{%
1+2\omega }\right) \right] =0,\right.   \nonumber
\end{eqnarray}
where \eqref{eq30} and \eqref{eq31} have also been used. 

Here, we will show two classes of solutions, the first one for $N=1$ and the second one for $N=2$. In this way it is possible to have exact results for $\rho(r)$ and $\psi(r)$ for each of the classes.

For $N=1$, we have

\begin{equation}
\rho(r)=\frac{-rF_1\pm\sqrt{r^2F_1^2+F_2(2H_1+r^2H_2)}}{rF_2}\sigma, \label{rhop1}
\end{equation}
with

\begin{eqnarray}
F_1=(1+2\beta)\sigma \left[7+2\omega-6\omega^2+3\kappa^2\omega(1+2\omega)+\kappa \beta(2+\omega)(1+2\omega)\right], \\
F_2=\kappa \sigma(1+2\beta)(1+2\omega)\left[2+\omega+3\kappa \omega+2\beta(2+\omega) \right], \\
H_1=2+\omega(7+6\omega)+2\beta \left[2+\omega(7+6\omega)\right], \\
H_2=\kappa \beta^2 \sigma \left\lbrace -2+\omega \left[-5-2\omega+18\kappa(1+\omega) \right]\right\rbrace. 
\end{eqnarray}

Hence, $\psi(r)$ is best described by

\begin{eqnarray}
\mbox{\footnotesize $ \psi(r)=\pm C_3r\sqrt{\frac{1}{3r^2}-\left[\frac{rF_1\mp\sqrt{r^2F_1^2+F_2(2H_1+r^2H_2)}}{3rF_2}\right]\sigma
+\frac{H_3}{6}\left\lbrace H_4-\left[\frac{rF_1\mp\sqrt{r^2F_1^2+F_2(2H_1+r^2H_2)}}{rF_2}\right]\sigma\right\rbrace ^2 }$}, \label{psi11}
\end{eqnarray}

with

\begin{eqnarray}
H_3=\frac{\kappa (1+2\beta)}{\sigma}, \\
H_4=\frac{\beta \sigma}{(1+2\beta)}.
\end{eqnarray}

In a similar manner, for $N=2$ we have

\begin{equation}
\rho(r)=\frac{-rG_1\pm\sqrt{r^2G_1^2+G_2(2H_1+r^2H_2)}}{rG_2}\sigma, \label{rhop2}
\end{equation}
with

\begin{eqnarray}
G_1=\kappa \sigma(1+2\beta)(1+2\omega)\left[3\kappa \omega+\beta(2+\omega) \right], \\
G_2=(1+2\beta)\sigma \left[3\kappa^2 \omega (1+2\omega)+\kappa (1+2\beta)(2+\omega)(1+2\omega)+2\sigma(7+2\omega-6\omega^2) \right].
\end{eqnarray}

Thereby, the $\psi(r)$ solutions for this class are

\begin{equation}
\mbox{\footnotesize $
\psi(r)=\pm C_3r \sqrt{\frac{1}{3r^2} -\left[\frac{rG_1\mp\sqrt{r^2G_1^2+G_2(2H_1+r^2H_2)}}{3rG_2
}\sigma \right]^2 +\frac{H_3}{6}\left\lbrace H_4 -\left[\frac{rG_1\mp\sqrt{r^2G_1^2+G_2(2H_1+r^2H_2)}}{rG_2} \right]\sigma \right\rbrace}.$} \label{psi21}
\end{equation}

\section{Energy Conditions Application}\label{sec:ec}

In this section we will construct the null energy condition (NEC) and weak energy condition (WEC) for the WH solutions previously obtained. The NEC and WEC read \cite{morris/1988}

\begin{eqnarray}
\rho^{\text{eff}} + p_t^{\text{eff}} \geq 0, \label{NEC} \\
\rho^{\text{eff}} \geq 0. \label{WEC}
\end{eqnarray}

Both $\rho ^{\text{eff}}  $ and $p_t ^{\text{eff}}$ in the energy conditions are expressed by Eqs.\eqref{eff} and  \eqref{eq16}, respectively, and, more important, are directly dependent of the solutions of $\rho (r)$ obtained above. If these conditions are respected for braneworld WHs, one may understand that their matter-energy content has non-negative values for any observer.

\begin{figure}
\vspace{0.3cm}
    \centering
    \includegraphics[height=4.8cm,angle=00]{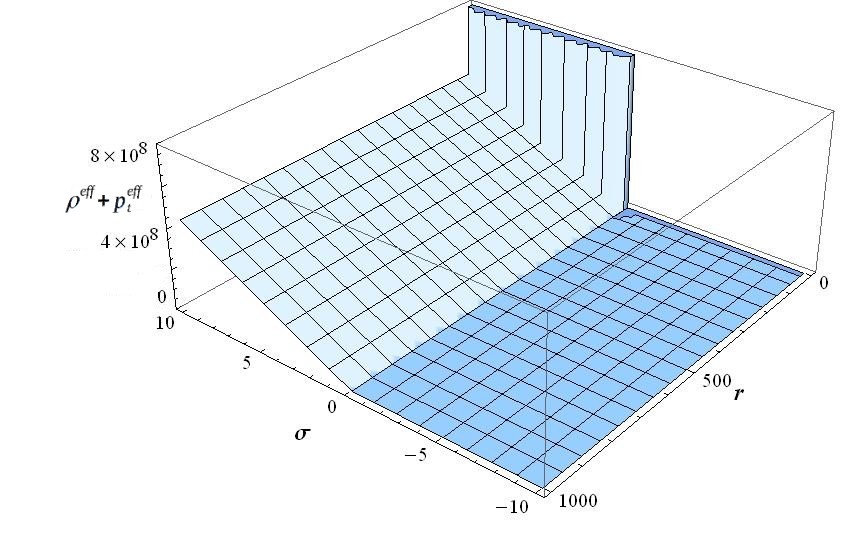}
    \includegraphics[height=4.8cm,angle=00]{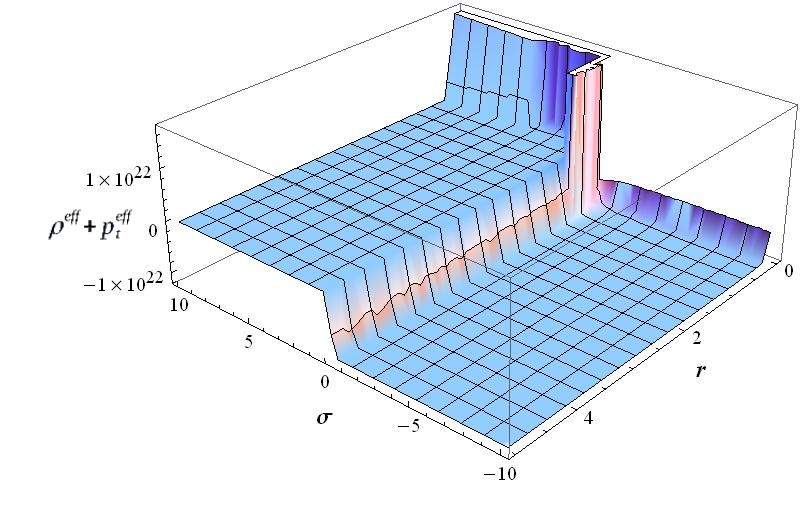}
    \caption{NEC from negative (left figure) and positive signs (right figure) in Eq.(\ref{rhop1})}
    \label{NEC1n}
\end{figure}

\begin{figure}
\vspace{0.3cm}
    \centering
    \includegraphics[height=4.9cm,angle=00]{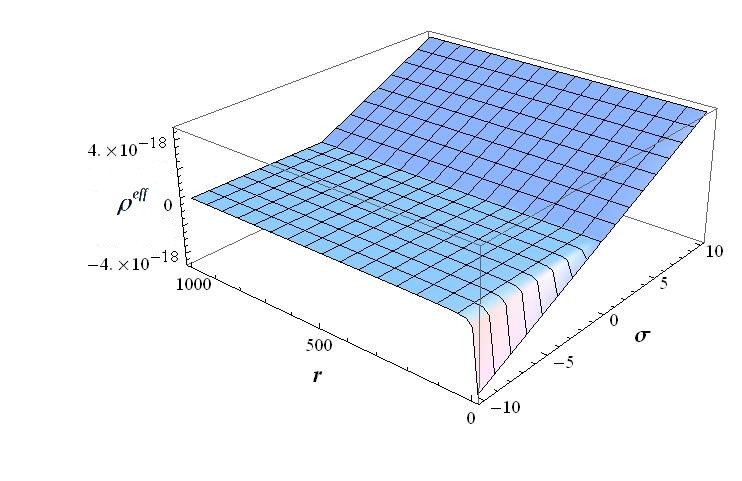}
     \includegraphics[height=4.9cm,angle=00]{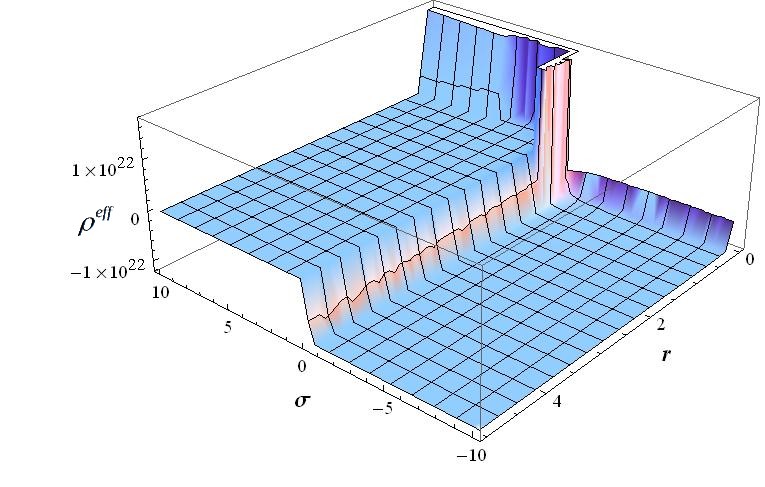}
    \caption{WEC from negative (left figure) and positive (right figure) signs in Eq.(\ref{rhop1})}
    \label{WEC1n}
\end{figure}

In Fig.\ref{NEC1n} we plot the NEC from the negative and positive signs in Eq.\eqref{rhop1}. In Fig.\ref{WEC1n}, the WEC is plotted also for both signs in \eqref{rhop1}. For all these cases we assume $N=1$.   

In the next section we will discuss the relevance of those results.







\section{Conclusions and final remarks}\label{sec:frc}

Braneworld models are important extended gravity scenarios nowadays and have been constantly put in test against observations \cite{abdujabbarov/2010,liddle/2003,keeton/2006,lombriser/2009,iorio/2005}. They have been used to evade some important issues in cosmology, such as the cosmological constant and dark matter problems \cite{dvali/2003,koyama/2008,diakonos/2009,shahidi/2011,matos/2004}.

Here, instead, we have used the RS II braneworld model as the underlying gravity scenario to construct static WH solutions. We assumed that the interior space-time of these objects admits conformal motion and from some hypothesis for the density-pressure relation inside them, we were able to obtain solutions regarding the WH material content.

In possession of the material solutions we constructed the energy conditions. Our intention was to check if the extra degrees of freedom provided by RS II braneworld scenario could allow the WHs in such a theory to be devoid of exotic matter. 

Indeed, Fig.\ref{NEC1n} have shown the possibility of respecting the energy conditions when $\sigma>0$. This is a quite interesting feature of the model, since it agrees with some important references that point to the gravitational instability of negative tension branes \cite{marolf/2001,charmousis/2004}.

\subsection*{Acknowledgments}
 PHRSM would like to thank S\~ao Paulo Research Foundation  (FAPESP), grant 2018/20689-7, for financial support. RACC thanks to 
FAPESP, grant numbers 2016/03276-5 and 2017/26646-5 for financial support. GR would like to thank the Brazilian National Research Council (CNPq) for providing scholarship program.

\end{document}